\documentclass[aps,prc,letterpaper,11pt,twoside,nofootinbib,showpacs,preprint]{revtex4}
\usepackage{graphicx}
\usepackage[sort&compress]{natbib}
\newcommand{\be}{\begin{equation}}
\newcommand{\ee}{\end{equation}}
\newcommand{\bea}{\begin{eqnarray}}
\newcommand{\eea}{\end{eqnarray}}
\newcommand{\bef}{\begin{figure}}
\newcommand{\eef}{\end{figure}}
\newcommand{\bce}{\begin{center}}
\newcommand{\ece}{\end{center}}

\begin{document}

\title{$\bar K^*$ meson in dense matter}

\author{L. Tol\'os$^1$, R. Molina$^2$, E. Oset$^2$ and A. Ramos$^3$\\
$^1$ Theory Group. KVI. University of Groningen, \\
Zernikelaan 25, 9747 AA Groningen, The Netherlands \\
$^2$ Instituto de F{\'\i}sica Corpuscular (centro mixto CSIC-UV)\\
Institutos de Investigaci\'on de Paterna, Aptdo. 22085, 46071, Valencia, Spain \\
$^3$ Estructura i Constituents de la Mat\`eria. Facultat de F\'{\i}sica. Universitat de Barcelona,\\
Avda. Diagonal 647, 08028 Barcelona, Spain
}

\begin{abstract}

We study the properties of $\bar K^*$ mesons in nuclear matter using a unitary
approach in coupled channels within the framework of the local hidden gauge
formalism and incorporating the $\bar K  \pi$ decay channel in matter. The
in-medium $\bar K^* N$ interaction accounts for Pauli blocking effects and
incorporates the $\bar K^*$ self-energy in a self-consistent manner.  We also
obtain the $\bar K^*$ (off-shell) spectral function and analyze its behaviour
at finite density and momentum. At normal nuclear matter density, the
$\bar K^*$ meson feels a moderately attractive potential while the $\bar K^*$
width becomes five times larger than in free space. We estimate the
transparency ratio of the $\gamma A \to K^+ K^{* -} A^\prime$ reaction, which
we propose as a feasible scenario at present facilities to detect the changes
of the properties of the $\bar K^*$ meson in the nuclear medium.

\end{abstract}


\vspace{1cm}

\date{\today}

\maketitle

\section{Introduction}
 \label{sec:intro}

The interaction of vector mesons with
nuclear matter has been the focus of attention for years and has been tied to
fundamental aspects of QCD \cite{rapp,Hayano:2008vn,Leupold:2009kz}. Yet, the theoretical
models offer a large variety of results from a large attraction to a large
repulsion. Early results on this issue within the Nambu Jona Lasinio model
produced no shift of the masses \cite{Bernard:1988db}  while, using qualitative
arguments, a universal large attractive shift of the mass was suggested in
\cite{Brown:1991kk}. More recent detailed calculations show no shift of the
mass of the $\rho$ meson in matter
\cite{Rapp:1997fs,Peters:1997va,Urban:1999im,Cabrera:2000dx,Post:2003hu} and very small for the $\phi$
meson \cite{Cabrera:2002hc}. Experimentally the situation has undergone big
steps recently, an example being the NA60 collaboration that reported a null shift
of the $\rho$ mass in the medium
\cite{Arnaldi:2006jq,Damjanovic:2007qm,Arnaldi:2008er} in the dilepton spectra
of heavy ion reactions. Detailed theoretical analyses of the reaction are also
done in  \cite{vanHees:2007th,Rapp:1999us} confirming those conclusions and
leaving room to some widening of the resonance. Similarly, a null shift in the
$\gamma$ induced dilepton production at CLAS is also found in
\cite{Wood:2008ee,Djalali:2008zza}. On the other hand, the KEK team had earlier reported  an
attractive mass shift of the $\rho$ in \cite{Muto:2005za,Naruki:2005kd}.  As
explained in detail in \cite{Wood:2008ee}, the different conclusions can be
traced back to the way the background is subtracted.

  The case of the $\omega$ in the medium has been more controversial.
Theoretically, there are about twenty different works claiming a variety of
mass shifts, from  largely attractive to largely repulsive (see
\cite{Caillon:1995ci,Saito:1998ev,Lutz:2001mi,Muehlich:2006nn} and \cite{Kaskulov:2006zc,Kaskulov:2006fi} for details).
Experimentally, a large shift of the mass of the $\omega$ was reported  from
the study of the photon induced $\omega$ production in nuclei, with the
$\omega$ being detected through its $\pi^0 \gamma$ decay channel \cite{prl}.
However, it was shown in \cite{Kaskulov:2006zc} that the shift could just be a
consequence of a particular choice of background subtraction and that other
reasonable  choices led to different conclusions. In fact, a recent reanalysis
of the background of the reaction has concluded that one cannot determine the
shift of the $\omega$ mass in the nucleus from that reaction \cite{Nanova:2010sy}. On the other
hand, the determination of the $\omega$ width in the medium by means of
the production cross section
in different nuclei, which leads to the transparency ratio
\cite{Hernandez:1992rv},
was rather
successful, giving rise to an appreciable in-medium enhancement \cite{Kaskulov:2006zc,:2008xy}.

  Curiously, no discussion has been made about the properties of the strange
vector mesons in the medium and this is the purpose of the present paper. The reason
might be that no much attention has been payed to the interaction of $K^*$ and
$\bar{K}^*$ mesons
with baryons. The fact that the $\bar{K}^*$ cannot be detected with dileptons might also have been a reason for the experimental neglect. The situation has been reversed only recently in Ref.~\cite{GarciaRecio:2005hy}, implementing the interaction light quark - light quark spin independent and SU(3) flavor independent, and in Ref.~\cite{Sarkar:2009kx,Oset:2009vf}, where this problem has been tackled using
the hidden local gauge formalism for the interaction of vector mesons with
baryons of the octet and the decuplet. The study of \cite{Oset:2009vf} allows
us to obtain the $\bar{K}^*$ self-energy in the nuclear medium,
following a similar approach to the one employed in \cite{Ramos:1999ku} for the
interaction of $\bar{K}$ in the medium starting from the $\bar{K} N$
interaction of the chiral unitary approach \cite{angels}.  We shall find that,
while the $\bar{K}^*$ feels a moderately attractive optical potential
at normal nuclear matter density,
there is a spectacular enhancement of the
$\bar{K}^*$ width in the medium, up to about five times the free value of
about 50 MeV.
      We also perform a qualitative estimate of the transparency ratio for
$\bar{K}^*$ production in the $\gamma~A \to K^+ \bar{K}^{*-}A'$ reaction in
order to motivate the experimental search of this remarkable medium property.

      The paper is organized as follows: the formalism is presented in Sections II and III, where we first review how the $\bar{K}^*N$ interaction in the free space is obtained and, next, we describe
the calculation of the $\bar{K}^*$ self-energy in nuclear matter, with
contributions coming from the $\bar{K}^* N$ absorption mechanism as well as from  $\bar{K}^*$ decay into
$\bar{K}\pi$ in dense matter. In Section IV we show the results obtained for the $\bar{K}^*$ properties in the
medium while in Section V we make an overview of the method including a critical discussion and future developments. Section VI is devoted to present the corresponding nuclear transparency ratio in the $\gamma
A \to K^*K^{*-}A'$ reaction and  our conclusions and outlook are given in Section VII.

\section{${\boldmath \bar K^* N}$ interaction in free space.}
\label{freespace}

We present in this section a review of the formalism used for building up the
interaction of vector mesons with baryons in free space. Medium modifications are
incorporated in the next section to obtain the $\bar K^*$ self-energy in nuclear matter.

In order to construct the interaction of vector mesons with baryons, we first obtain the interaction between vector mesons. We follow the formalism of the hidden gauge interaction for vector mesons of
\cite{hidden1,hidden2,hidden3,hidden4} (see also \cite{hidekoroca} for a practical set of Feynman rules).
The  Lagrangian involving the interaction of
vector mesons amongst themselves is given by
\begin{equation}
{\cal L}_{III}=-\frac{1}{4}\langle V_{\mu \nu}V^{\mu\nu}\rangle \ ,
\label{lVV}
\end{equation}
where the symbol $\langle \rangle$ stands for the trace in the SU(3) space
and $V_{\mu\nu}$ is given by
\begin{equation}
V_{\mu\nu}=\partial_{\mu} V_\nu -\partial_\nu V_\mu -ig[V_\mu,V_\nu]\ ,
\label{Vmunu}
\end{equation}
where  $g$ is
\begin{equation}
g=\frac{M_V}{2f}\ ,
\label{g}
\end{equation}
with $M_V$ being the vector meson mass
and $f=93$ MeV the pion decay constant. The value of $g$ given by the
relation of Eq. (\ref{g})
fulfills the KSFR rule \cite{KSFR} which is tied to
vector meson dominance \cite{sakurai}. The magnitude $V_\mu$ is the SU(3)
matrix of the vectors of the octet of the $\rho$ plus the SU(3) singlet
\begin{equation}
V_\mu=\left(
\begin{array}{ccc}
\frac{\rho^0}{\sqrt{2}}+\frac{\omega}{\sqrt{2}}&\rho^+& K^{*+}\\
\rho^-& -\frac{\rho^0}{\sqrt{2}}+\frac{\omega}{\sqrt{2}}&K^{*0}\\
K^{*-}& \bar{K}^{*0}&\phi\\
\end{array}
\right)_\mu \ .
\label{Vmu}
\end{equation}
From this Lagrangian we obtain a three-vector vertex term
\begin{equation}
{\cal L}^{(3V)}_{III}=ig\langle (V^\mu\partial_\nu V_\mu -\partial_\nu V_\mu
V^\mu) V^\nu\rangle
\label{l3V}\ ,
\end{equation}
analogous to the coupling of vectors to pseudoscalars.
In a similar way, one obtains the Lagrangian for the coupling of vector mesons to
the baryon octet given by
\cite{Klingl:1997kf,Palomar:2002hk}:
\begin{equation}
{\cal L}_{BBV} = g\left( \langle \bar{B}\gamma_{\mu}[V^{\mu},B]\rangle +
\langle \bar{B}\gamma_{\mu}B \rangle \langle V^{\mu}\rangle \right) ,
\label{lagr82}
\end{equation}
where $B$ is now the SU(3) matrix of the baryon octet
\begin{equation}
B =
\left(
\begin{array}{ccc}
\frac{1}{\sqrt{2}} \Sigma^0 + \frac{1}{\sqrt{6}} \Lambda &
\Sigma^+ & p \\
\Sigma^- & - \frac{1}{\sqrt{2}} \Sigma^0 + \frac{1}{\sqrt{6}} \Lambda & n \\
\Xi^- & \Xi^0 & - \frac{2}{\sqrt{6}} \Lambda
\end{array}
\right) \ .
\end{equation}

With these ingredients we can construct the Feynman diagrams that lead to  the vector-baryon ($VB$) transitions $VB
\to V^\prime B^\prime$. As discussed in Ref.~\cite{Oset:2009vf}, one can proceed
by neglecting the three momentum of the external vectors versus the vector
mass, in a similar way as done for chiral Lagrangians under the low energy
approximation, and one obtains the transition potential:
\begin{equation}
V_{i j}= - C_{i j} \, \frac{1}{4 f^2} \, \left( k^0 + k^\prime{}^0\right)
~\vec{\epsilon}\,\vec{\epsilon }\,^\prime , \label{kernel}
\end{equation}
where
$k^0, k^\prime{}^0$ are the energies of the incoming and outgoing vector
mesons, respectively, $\vec{\epsilon}\,\vec{\epsilon }\,^\prime$ is the product of their
polarization vectors, and $C_{ij}$ are the symmetry coefficients
\cite{Oset:2009vf}.

The meson-baryon scattering amplitude is obtained from the coupled-channel
on-shell Bethe-Salpeter equation \cite{angels,ollerulf}
\begin{equation}
T = [1 - V \, G]^{-1}\, V \ ,
\label{eq:Bethe}
\end{equation}
with $G$ being the loop
function of a vector meson of mass $m$ and a baryon of mass $M$ with total
four-momentum $P$ ($s=P^2$):
\begin{eqnarray} G(s,m^2,M^2) &=& {\rm i} 2 M
\int \frac{d^4 q}{(2 \pi)^4} \, \frac{1}{(P-q)^2 - M^2 + {\rm i} \epsilon} \,
\frac{1}{q^2 - m^2 + {\rm i} \epsilon}   \ ,
%
\label{eq:gpropdr}
\end{eqnarray}
which is conveniently regularized
\cite{ollerulf} taking a natural value of $-2$ for the subtraction constants
at a regularization scale $\mu=630$ MeV \cite{Oset:2009vf}.

We note that the iteration of diagrams implicit in the Bethe-Salpeter equation
for vector meson scattering implies a sum over the polarizations of the
internal vector mesons which, because they are tied to the external ones
through the $\vec{\epsilon}\,\vec{\epsilon }\,^\prime$ factor, leads to a
correction in the $G$ function of $\vec{q}\,^{2}/3 M_V^2$ \cite{luisaxial}, a
factor that can be safely neglected in consonance with the low-momentum approximation
done. This leads to the factorization  of  the factor
$\vec{\epsilon}\,\vec{\epsilon }\,^\prime$ for the
external vector mesons also in the $T$ matrix.
This method provides degenerate pairs of
resonances which have  $J^P=1/2^-,3/2^-$, a pattern that seems to be reproduced by
the existing experimental data (see in the PDG \cite{Amsler:2008zzb} the states
$N^*(1650) (1/2^-)$, $N^*(1700) (3/2^-)$, $N^*(2080) (3/2^-)$, $N^*(2090) (1/2^-)$, $\Sigma(1940) (3/2^-)$, $\Sigma(2000) (1/2^-)$, $\Delta(1900) (1/2^-)$, $\Delta(1940)(3/2^-)$, $\Delta(1930)(5/2^-)$).

Since we are interested in studying the interaction of $\bar K^*$ mesons in
nuclear matter, we concentrate in the strangeness $S=-1$ vector meson-baryon
sector with isospin $I=0$ and $I=1$. For $S=-1,I=0$ we find five vector
meson-baryon coupled channels ($\bar K^*N$, $\omega \Lambda$, $\rho \Sigma$,
$\phi \Lambda$ and $K^* \Xi$), while for $S=-1,I=1$ we consider six channels
($\bar K^*N$, $\rho \Lambda$, $\rho \Sigma$, $\omega \Sigma$, $K^* \Xi$ and
$\phi \Sigma$).

The relatively large decay width of the $\rho$ and  $\bar K^*$ vector mesons
(into $\pi\pi$ or $\bar K\pi$ pairs, respectively)
is incorporated in the loop functions via the
convolution \cite{nagahiro}:
\begin{eqnarray}
\tilde{G}(s)= \frac{1}{N}\int^{(m+\Delta_r)^2}_{(m-\Delta_l)^2}d\tilde{m}^2
\left(-\frac{1}{\pi}\right)
{\rm Im}\,\frac{1}{\tilde{m}^2-m^2+{\rm i} m \Gamma(\tilde{m})}
& G(s,\tilde{m}^2,M^2)\ ,
\label{Gconvolution}
\end{eqnarray}
with
\begin{equation}
N=\int^{(m+\Delta_r)^2}_{(m-\Delta_l)^2}d\tilde{m}^2
\left(-\frac{1}{\pi}\right){\rm Im}\,\frac{1}{\tilde{m}^2-m^2+{\rm i}m
\Gamma(\tilde{m})}
\label{Norm}
\end{equation}
being the normalization factor. The integration range around the meson mass is
established by the left and right parameters $\Delta_l$, $\Delta_r$,
taken to be a few
times the meson width $\Gamma$, the value of which is $149.4$ MeV for the $\rho$
 meson and $50.5$ MeV  for the
$\bar K^*$ meson.
The energy dependent width $\Gamma(\tilde{m})$ is obtained from
\begin{equation}
\Gamma(\tilde{m})=\Gamma \frac{m^2}{\tilde{m}^2}
\frac{q(\tilde{m})^3}{q(m)^3} \theta(\tilde{m}-m_1-m_2) \ ,
\label{gamma}
\end{equation}
where $q(\sqrt{s})$ is the momentum of the decay products in the rest frame
of the vector meson with invariant mass $\sqrt{s}$. In the case of
$\rho$ decay we have $m_1=m_2=m_\pi$, while for
$\bar K^*$ decay $m_1=m_\pi$, $m_2=m_{\bar K}$.

\section{${\boldmath \bar K^*}$ self-energy in nuclear matter}

There are two sources for the modification of the $\bar K^*$ self-energy in
nuclear matter: a) the contribution associated to the decay mode $\bar K \pi$
modified by the nuclear medium effects on the $\bar K$ and $\pi$ mesons, and b)
the contribution associated to the interaction of the $\bar K^*$ with the
nucleons in the medium, which accounts for the direct quasi-elastic process
$\bar K^* N \to \bar K^* N$ as well as other absorption channels  $\bar K^*
N\to \rho Y, \omega Y, \phi Y, \dots$ with $Y=\Lambda,\Sigma$.

\subsection{${\boldmath \bar{K}^*}$ self-energy from decay into ${\boldmath \bar{K}\pi}$}

\begin{figure}[t]
\begin{center}
\includegraphics[width=0.8\textwidth]{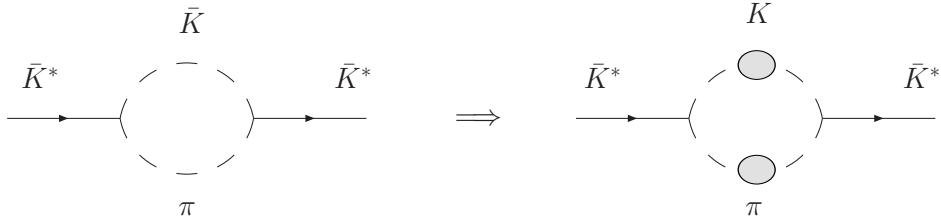}
\caption{The $\bar{K}$ propagator renormalized to allow its decay into $\bar{K}\pi$, in the free space (left), and in  the medium (right), including the self-energies of the $\bar{K}$ and $\pi$ mesons. }
\label{fig:1}
\end{center}
\end{figure}

We devote this section to study the decay of the $\bar{K}^*$ meson into $\bar{K}\pi$ pairs renormalized in the medium.
 First of all, to test our model, we evaluate the theoretical width in the vacuum, indicated by the diagram on the left in Fig.~\ref{fig:1},
  including the $K^{*-} \to \bar K^0 \pi^-$ and $K^{*-} \to \bar K^- \pi^0$ processes, which gives:
 \begin{equation}
\Pi_{\bar{K}^*}^{0}(q^0,\vec{q}\,)=2 g^2\vec{\epsilon}\cdot \vec{\epsilon}\,'\int
\frac{d^4 k}{(2\pi)^4}  \frac{\vec{k}\,^2}{k^2-m_\pi^2}\frac{1}{(q-k)^2-m_{\bar K}^2+i\epsilon}\ ,
\label{eq:freeint}
\end{equation}
where the $VPP$ vertices have been obtained from the hidden gauge lagrangian of
${\cal L}_{VPP}=-ig\langle V^\mu [\phi,\partial_\mu \phi] \rangle$ \cite{hidden1,hidekoroca},
within the low momentum approximation employed in the present work. This approximation  has allowed us to
keep only the spatial components of the vector polarizations, that means, substituting
$\epsilon_\mu k^\mu \epsilon'_\nu k^\nu$ by $\epsilon_i k_i \epsilon'_j k_j$, a factor
that has been further replaced in the integral by $\vec{\epsilon} \cdot \vec{\epsilon}\,' \,\frac{1}{3} \vec{k}\,^2\, \delta_{ij}$.

The imaginary part of the free $\bar K^*$ self-energy at rest:
\begin{equation}
\mathrm{Im}\Pi_{\bar{K}^*}^{0}(q^0=m_{\bar K^*},\vec{q}=0)=\frac{g^2}{4\pi} \vec{\epsilon}\cdot\vec{\epsilon}\,'\,\mid {\vec k} \mid^3\frac{1}{m_{\bar K^*}}\ ,
\end{equation}
where the momentum of the emitted mesons is $\vert \vec{k} \vert = 288.7$ MeV, determines a value of the $K^{*-}$ width  of $\Gamma_{K^{*-}}=-\mathrm{Im}\Pi_{\bar{K}^*}^{0}/m_{\bar K^*}=42$ MeV, which is quite close to the experimental value $\Gamma^{\rm exp}_{K^{*-}}=50.8\pm 0.9$ MeV.


In the medium we shall calculate the $\bar{K}^*$ self-energy coming from its
decay into ${\bar K}\pi$ including both the self-energy of the antikaon and the pion. This
will add new contributions to the $\bar K^*$ self-energy as the one depicted by
the second diagram in Fig. \ref{fig:1}.


For completeness, we briefly recall the essential features of the models employed for
the properties of antikaons and pions in nuclear matter. We refer to \cite{Ramos:1999ku,Tolos:2006ny,Oset:1989ey,Ramos:1994xy,Waas:1996fy,Waas:1997pe,GarciaRecio:2002cu} for a detailed explanation.

The $\bar{K}$ self-energy in symmetric nuclear matter is obtained from the antikaon-nucleon interaction within a chiral unitary approach. The model incorporates $s$- and $p$-waves in the interaction and includes the following channels: $\bar{K}N$, $\pi\Sigma$, $\eta\Lambda$, $K\Xi$ for $I=0$, and $\bar{K}N$, $\pi\Lambda$, $\pi\Sigma$, $\eta\Sigma$, $K\Xi$ for $I=1$. The tree level $s$-wave amplitudes are given by the Weinberg-Tomozawa term of the lowest order chiral Lagrangian describing the interaction of the pseudoscalar meson octet with the $1/2^+$ baryon octet. Unitarization in coupled channels is imposed by solving the Bethe-Salpeter equation with on-shell amplitudes. Loops incorporate in-medium effects, which include Pauli-blocking corrections, as well as mean-field binding on the nucleons and hyperons via a $\sigma-\omega$ model. They are regularized with
a cutoff momentum of $q_{\rm max}=630$ MeV.
 Self-consistency in the $\bar{K}$ self-energy is also required.
 The $p$-wave contribution to the $\bar{K}$ self-energy is built up from the coupling of the $\bar{K}$ meson to $\Lambda N^{-1}$, $\Sigma N^{-1}$ and $\Sigma^* N^{-1}$ excitations.
The model determines an attractive optical potential of $U_{K^-}=\mathrm{Re}\Pi_{K^-}/2 m_{K^-}\sim -66$ MeV. This attraction is quite moderate in comparison with other approximation schemes \cite{Li:1997zb,Li:1997tz,Mao:1998nv,Schaffner:1995th,Schaffner:1996kv,Tsushima:1997df},
although it agrees with others which also implement self-consistency
\cite{Lutz1,Schafner,Cieply,Tolos:2000fj,Tolos:2002ud}.

The pion self-energy consists of a small $s$-wave part, $\Pi^{(s)}_\pi(\rho)$ (momentum independent), and a dominant $p$-wave component, $\Pi^{(p)}_\pi(k^0,\vec{k},\rho)$, that comes from particle-hole ($N N^{-1}$ or $ph$)  and $\Delta N^{-1}$
excitations \cite{Oset:1989ey}. The model also includes a two-particle-two-hole ($2p2h$) piece tied to two-nucleon pion absorption.
The strength of the considered collective modes is modified by repulsive, spin-isospin $NN$ and $N\Delta$ short-range correlations, which are included in a phenomenological way with a single Landau-Migdal interaction parameter, $g'$ \cite{Ramos:1994xy}.

Replacing the pion and antikaon propagators in Eq.~(\ref{eq:freeint}) by their respective propagators in the medium, written in the Lehmann representation, gives:
\begin{eqnarray}
-i\Pi_{\bar{K}^*}^{\rho,{\rm (a)}}(q^0,\vec{q}\,)&= &2\,g^2\vec{\epsilon}\cdot \vec{\epsilon}\,'\int \frac{d^4 k}{(2\pi)^4} \vec{k}\,^2 \int^\infty_0\frac{d\omega}{\pi}(-2\omega)\frac{\mathrm{Im}D_\pi(\omega,\vec{k}\,)}
{(k^{0})^2-\omega^2+i\epsilon}\nonumber\\&\times&\int^\infty_0\frac{d\omega'}
{\pi}(-)
\left\{\frac{\mathrm{Im} D_{\bar{K}}(\omega',\vec{q}-\vec{k}\,)}{q^0-k^0-\omega'+i\eta}-\frac{\mathrm{Im} D_K(\omega',\vec{q}-\vec{k}\,)}{q^0-k^0+\omega'-i\eta}\right\}\ ,
\end{eqnarray}
which, after integrating over the $k^0$ variable, becomes
\begin{eqnarray}
\Pi_{\bar{K}^*}^{\rho,{\rm (a)}}(q^0,\vec{q}\,)&=&2g^2\vec{\epsilon}\cdot \vec{\epsilon}\,'\int \frac{d^3 k}{(2\pi)^3}\frac{\vec{k}\,^2}{\pi^2}\int^\infty_0 d\omega\mathrm{Im} D_\pi(\omega,\vec{k}\,)  \nonumber\\&\times&
\int^\infty_0 d\omega' \left\{ \frac{\mathrm{Im}D_{\bar{K}}(\omega',\vec{q}-\vec{k}\,)}{q^0-\omega-\omega'+i \eta}-\frac{\mathrm{Im} D_K(\omega',\vec{q}-\vec{k}\,)}{q^0+\omega+\omega'-i\eta}\right\}\ .
\label{eq:kaon}
\end{eqnarray}
Since we are using the physical mass of the $\bar K^*$, the real part of this in-medium self-energy must vanish at $\rho=0$.
This is achieved by
subtracting the real part of the free $\bar{K}^*$ self-energy, $\Pi_{\bar{K}^*}^{0}$, to $\Pi^{\rho,{\rm (a)}}_{\bar{K}^*}$.
The part of negative energy of the $\bar{K}$ propagator, represented by the last term in Eq.~(\ref{eq:kaon}) is small, it does not contribute to the imaginary part of the $\bar K^*$ self-energy and, to a very good approximation, it can be canceled with the corresponding term of the free $\bar{K}^*$ self-energy.
Hence, we can write the $\bar{K}^*$ self-energy as
\begin{eqnarray}
\Pi_{\bar{K}^*}^{\rho,{\rm (a)}}(q^0,\vec{q}\,)&=&2g^2\vec{\epsilon}\cdot \vec{\epsilon}\,' \left\{ \int \frac{d^3 k}{(2\pi)^3} \frac{\vec{k}\,^2}{\pi^2}\int^\infty_0 d\omega \,\mathrm{Im} D_\pi( \omega,\vec{k}\,)\int^\infty_0 d\omega'\frac{\mathrm{Im} D_{\bar{K}}(\omega',\vec{q}-\vec{k}\,)}{q^0-\omega-\omega'+i\eta}\right. \nonumber\\
&-&\left. \mathrm{Re}\int \frac{d^3 k}{(2\pi)^3}\frac{\vec{k}\,^2}{2\omega_\pi(k)}
\frac{1}{2\omega_{\bar K}(q-k)}\frac{1}{q^0-\omega_\pi(k)-\omega_{\bar K}(q-k)+i\epsilon}\right\}\ .\label{eq:selfkpi}
\end{eqnarray}

\begin{figure}[t]
\begin{center}
\includegraphics[width=0.8\textwidth]{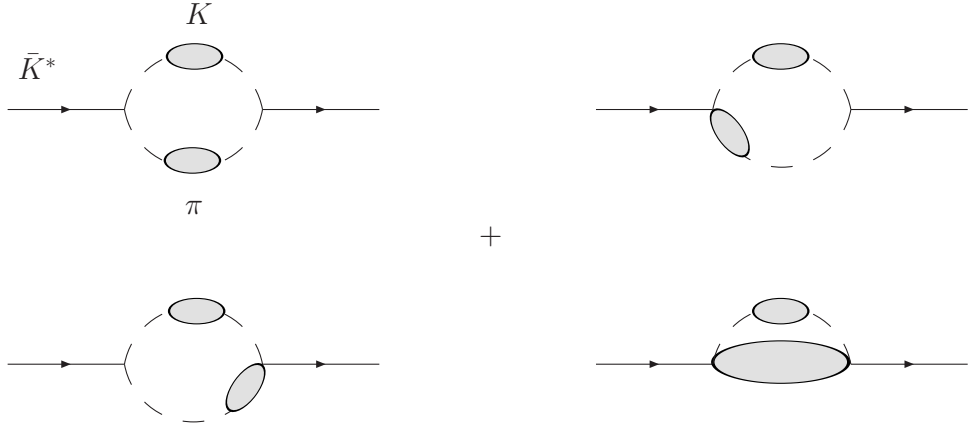}
\caption{Self-energy diagrams at first order in the nuclear density contributing to the decay of the $\bar{K}^*$ meson in the medium. }
\label{fig:3}
\end{center}
\end{figure}

One still has to implement vertex corrections which are required by the gauge invariance of the model \cite{Herrmann:1993za,Chanfray:1993ue}, and are associated to the last three diagrams of Fig. \ref{fig:3}. As we will see in the results section, the effect of dressing the pion is considerably larger than the inclusion of the $\bar{K}$ self-energy. Therefore, we consider the vertex corrections only for the case of the pion. They can be easily implemented by evaluating the first diagram of Fig.~\ref{fig:3} but
replacing the $p$-wave pion self-energy by
\begin{eqnarray}
\Pi_{\pi}^{(p)} \Longrightarrow
\frac{\Pi_{\pi}^{(p)}}{\vec{k}\,^2} \left( \vec{k}\,^2+\left[D_{\pi}^{0}(k)\right]^{-1}+
\frac{3}{4}\frac{\left[D_{\pi}^{0}(k)\right]^{-2}}{\vec{k}\,^2}
\right)\ ,
\end{eqnarray}
with $\left[D_{\pi}^{0}(k)\right]^{-1}=(k^0)^2-\vec{k}\,^2-m^2_\pi$ \cite{Oset:2000eg}.

\subsection{$\bar K^*$ self-energy from the s-wave $\bar K^* N$ interaction}

In this section we consider the contributions to the $\bar K^*$ self-energy
coming from its interactions with the nucleons in the Fermi sea. These are
implemented by incorporating the corresponding medium modifications in the
effective $\bar K^*N$ interaction. One of the sources of density dependence
comes from the Pauli principle acting on the nucleons, which prevents them from
being scattered into already occupied states. Another source is related to the
change of the properties of mesons and baryons in the coupled channel states
due to the interaction with nucleons of the Fermi sea. In particular, we
consider the self-consistently calculated $\bar K^*$ self-energy in the $\bar
K^*N$ intermediate states.

We proceed as done in Ref.~\cite{Tolos:2009nn}. Formally, the meson-baryon propagator in nuclear matter reads
\begin{eqnarray}
G^\rho(P) &=& G^0 (\sqrt{s})+ \mathrm{lim}_{\Lambda\to\infty}\delta G^\rho_\Lambda(P), \nonumber \\
\delta G^\rho_\Lambda(P) &\equiv& G^\rho_\Lambda(P)-G^0_\Lambda(\sqrt{s})~=~{\rm i}
2 M \int_\Lambda \frac{d^4 q}{(2 \pi)^4} \left ( D^\rho_{\cal
B}(P-q)~D^\rho_{\cal M}(q)-D^0_{\cal B}(P-q)~D^0_{\cal M}(q) \right ), \
\label{eq:deltaG}
\end{eqnarray}
where $\sqrt{s}$ is the center-of-mass energy and $q$ and $P-q$ the meson and baryon four-momentum, respectively.
The meson-baryon loop $G^0$ is calculated using dimensional regularization, while the medium $\delta G^\rho=\mathrm{lim}_{\Lambda\to\infty}\delta G^\rho_\Lambda(P)$ correction contains all the nuclear medium effects affecting the loop. This quantity is calculated in the infinite cutoff limit and, therefore, it is UV finite and independent of the selected subtracting point used to regularize $G^0$.

For the $\bar K^*N$ channel we consider Pauli blocking effects
on the nucleons together with self-energy insertions of the $\bar K^*$ meson. The self-energy is obtained self-consistently from the in-medium $\bar K^*N$ effective interaction,
${T^\rho}_{\bar K^*N}$, as we will show in the following. The
corresponding in-medium single-particle propagators are then given by
\bea
D^\rho_N(p)~&=& {1\over 2 E_N(\vec{p}\,)}
\left\{ \sum_r  u_r(\vec{p}\,) \bar{u}_r(\vec{p}\,)\left( \frac{1-n(\vec{p}\,)}{p^0 -E_N(\vec{p}\,) + i \varepsilon}+
\frac{n(\vec{p}\,)}{p^0 - E_N(\vec{p}\,)-i \varepsilon} \right) +
\frac{\sum_r v_r(-\vec{p}\,) \bar{v}_r(-\vec{p}\,)}{p^0+E_N(\vec{p}\,) - i \varepsilon} \right\} \nonumber \\
&=& D^0_N(p)~+~2\pi i~n(\vec{p}\,)~\frac{\delta\left (p^0-E_N({\vec p}\,)\right )}{2E_N(\vec{p}\,)}  \sum_r  u_r(\vec{p}\,) \bar{u}_r(\vec{p}\,)\ ,\\
D^\rho_{\bar K^*}(q)&=&\left ((q^0)^2 -\omega({\vec q}\,)^2-\Pi_{\bar K^*}(q) \right )^{-1}\,=\,
\int_{0}^{\infty} \, d\omega \, \left(
\frac{S_{\bar K^*}(\omega,\vec{q}\,)}{q^0-\omega +i
\varepsilon}-\frac{S_{K^*}(\omega,\vec{q}\,)}{q^0+\omega-i
\varepsilon} \right) ,~~
\label{eq:Krho}
\eea
where $E_N(\vec{p}\,)=\sqrt{{\vec{p}\,}^2+M_N^2}$ and $ \omega({\vec
q}\,)=\sqrt{{\vec{q}\,}^2+m_{\bar K^*}^2}$ are the nucleon and ${\bar K^*}$ energies, respectively,
$\Pi_{\bar K^*}(q^0,\vec{q}\,)$ is the $\bar K^*$ meson self-energy, and
$S_{\bar K^*}$ the corresponding meson spectral function. In a very good
approximation the spectral function for the $K^*$ meson  can be approximated
by the free-space one, viz. by a delta function, because there are no baryon resonances
in the strangeness $S=1$  case. Finally,
$n(\vec{p}\,)$ is the Fermi gas nucleon momentum distribution, given
by the step function $n(\vec{p}\,) = \Theta(p_F-|\vec{p}\,|)$, with
$p_F=(3\pi^2\rho/2)^{1/3}$.

Using Eq.~(\ref{eq:deltaG}) and performing the energy integral
over $q^0$, the $\bar K^*N$ loop function reads
\begin{eqnarray}
{G^\rho}_{\bar K^*N}(P)=
 G^0_{\bar K^*N}(\sqrt{s})+
\int \frac{d^3 q}{(2 \pi)^3} \,
\frac{ M_N }{ E_N(\vec{p}\,)} \,
\left [
\frac{-n(\vec{p}\,)}{(P^0 - E_N(\vec{p}\,))^2-\omega(\vec{q}\,)^2+i\varepsilon}
\, +\right.\phantom{largoooooooooo} &&
\label{eq:Glarga}\\
\left.
(1-n(\vec{p}\,))
\left. \left (
\frac{-1/(2 \omega({\vec q}\,))}
{P^0 -E_N(\vec{p}\,)-\omega(\vec{q}\,)+i \varepsilon}
+
\int_{0}^{\infty} \,
 d\omega \,
\frac{S_{\bar K^*}(\omega,\vec{q}\,)}{P^0 -E_N(\vec{p}\,)-\omega+i\varepsilon}
\, \right ) \right ]  \right| _{{\vec p}={\vec P}-{\vec q}}\ ,&&
\nonumber
\end{eqnarray}
where the first term of the integral, proportional to $-n(\vec{p}\,)$, is
the Pauli correction and accounts for the
case where the Pauli blocking on the nucleon is considered and
the meson in-medium self-energy is neglected.  The
second term, proportional to $(1-n(\vec{p}\,))$, would be exactly zero if
the meson spectral functions $S_{\bar K^*}$ was the free one,
$S^{0}_{\bar K^*}(\omega,\vec{q}\,)=
\delta(\omega- \omega(\vec{q}\,))/(2 \omega)$.
Then, it accounts for the contribution of the in-medium meson modification to
the loop function.

For all other meson-baryon states we use the free
loop functions given by Eq.~(\ref{eq:gpropdr}) (or Eq.~(\ref{Gconvolution})
for  $\rho$-baryon states) because
these meson-baryon states couple more moderately to
the $\bar K^* N$ channel.

We can now solve the on-shell Bethe-Salpeter equation in nuclear matter for
the in-medium amplitudes in the isospin basis
\begin{eqnarray}
{T^\rho}^{(I)}(P) &=& \frac{1}{1-
V^{I}(\sqrt{s})\,{G^\rho}^{(I)}(P)}\,V^{I}(\sqrt{s}) \ .
 \label{eq:scat-rho}
\end{eqnarray}
The in-medium $\bar K^*$ self-energy is  then obtained by
integrating ${T^\rho}_{\bar K^*N}$ over the nucleon Fermi sea,
\begin{eqnarray}
\Pi_{\bar{K}^*}^{\rho,{\rm (b)}}(q^0,\vec{q}\,)&=&\int \frac{d^3p}{(2\pi)^3} \, n(\vec{p}\,)\,
\left [~{T^\rho}^{(I=0)}_{\bar K^*N}(P^0,\vec{P})+3 {T^\rho}^{(I=1)}_{\bar K^*N}(P^0,\vec{P})\right ] \ ,
 \label{eq:pid}
\end{eqnarray}
where $P^0=q^0+E_N(\vec{p}\,)$ and $\vec{P}=\vec{q}+\vec{p}$ are the
total energy and momentum of the $\bar K^*N$ pair in the nuclear
matter rest frame, and the values $(q^0,\vec{q}\,)$ stand for the
energy and momentum of the $\bar K^*$ meson also in this frame. The
self-energy $\Pi_{\bar{K}^*}^{\rho,{\rm (b)}}(q^0,\vec{q}\,)$ has to be determined self-consistently
since it is obtained from the in-medium amplitude
${T}^\rho_{\bar K^*N}$ which contains the $\bar K^*N$ loop function
${G}^\rho_{\bar K^*N}$, and this last quantity itself is a function of
$\Pi_{\bar K^*}^{\rho}=\Pi_{\bar{K}^*}^{\rho,{\rm (a)}}
+\Pi_{\bar{K}^*}^{\rho,{\rm (b)}}$. From this self-energy we obtain the corresponding
spectral function which is used in the integral for the loop function
${G^\rho}_{\bar K^*N}(P^0, \vec{P} \,)$, as given
in Eq.~(\ref{eq:Glarga}).

\section{Results}

\begin{figure}[t]
\begin{center}
\includegraphics[height=0.5\textwidth,width=0.5\textwidth]{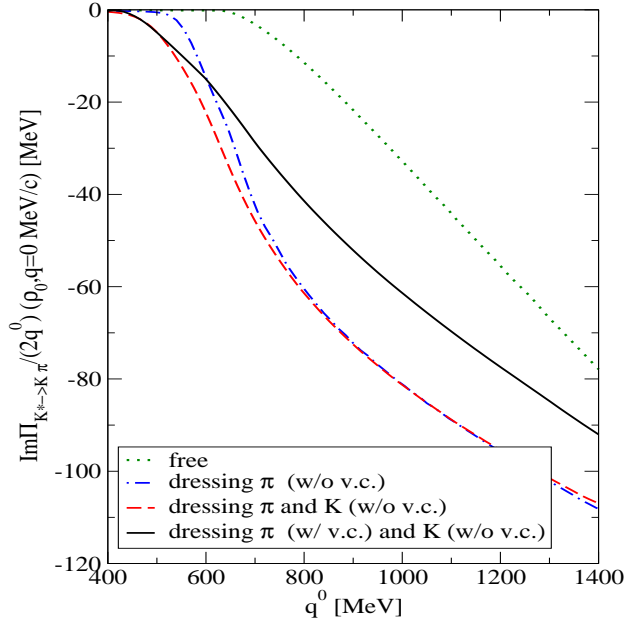}
\caption{Imaginary part of the $\bar K^*$ self-energy at zero momentum,
 coming from the ${\bar K}\pi$ decay mode in
dense matter at normal saturation density $\rho_0$.
Different approaches are studied: (i) calculation in free space, (ii) including the $\pi$ self-energy, (iii) including the $\pi$ and $\bar K$ self-energies, and (iv) including the $\pi$ dressing with vertex corrections and the $\bar K$ self-energy. }
\label{fig:kskpi}
\end{center}
\end{figure}

In Fig. \ref{fig:kskpi} we show the energy dependence of the imaginary part of
the $\bar{K}^*$ self-energy for $\vec{q}=0$ coming from $\bar{K}\pi$ decay, in
free space (dotted line),  adding the $\pi$ self-energy (dot-dashed line) at
normal nuclear matter saturation density ($\rho_0=0.17$ \ fm$^{-3}$)
and including both $\pi$ and
$\bar{K}$  self-energy contributions (dashed line). We can see in this figure
that including the self-energy of the pion changes considerably the in-medium
$\bar{K}^*$ width, which becomes, at normal nuclear matter density,
 about $3$ times larger than in vacuum. Including
the medium effects of the $\bar{K}$ has a very small effect. The reason for the
pion self-energy to have such a strong influence lies in the fact that the
$\bar K^*  \to \bar K \pi$ decay process leaves the pion with energy right in
the region of $\Delta N^{-1}$ excitations, where there is considerable pionic
strength. When the vertex corrections are included (continuous line in Fig.
\ref{fig:kskpi}) the effect of dressing the pion turns out to be more moderate,
giving a $\bar K^*$ width of
$\Gamma_{\bar{K}^*}(\rho=\rho_0)=105$ MeV, which is about twice the value of the width in vacuum.

\begin{figure}[t]
\begin{center}
\includegraphics[width=0.8\textwidth,]{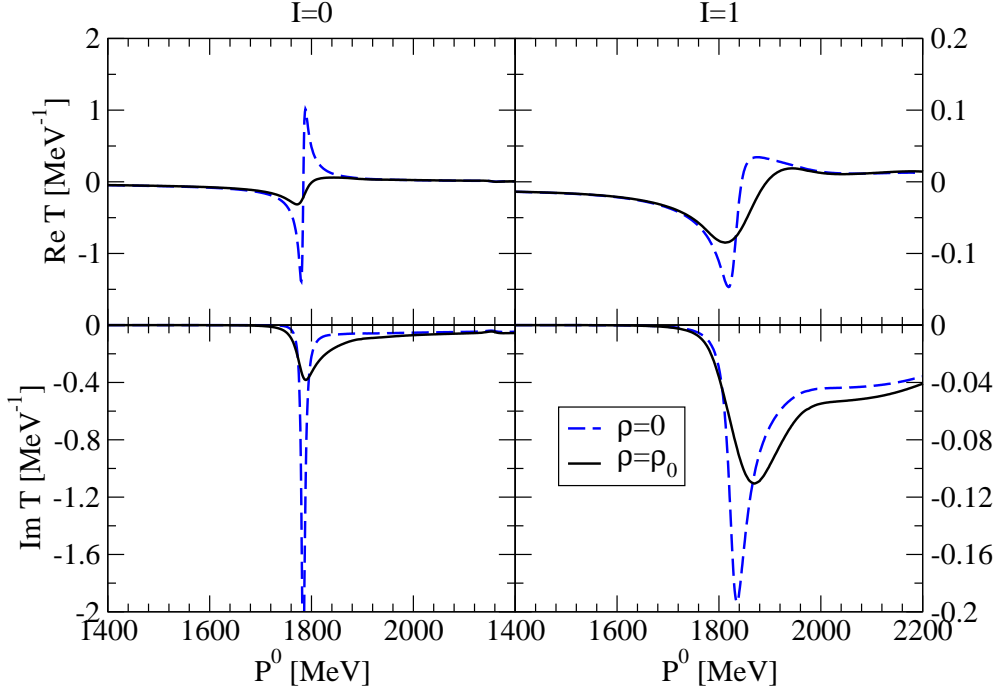}
\caption{Real and imaginary parts of the  $\bar K^*N \rightarrow \bar K^* N$ amplitude
as function of the center-of-mass energy $P_0$ for a fixed
total momentum $|\vec{P}|=0$. Two new states are generated dynamically: ($I=0$) $\Lambda(1783)$ and ($I=1$) $\Sigma(1830).$
}
\label{fig:reso}
\end{center}
\end{figure}

In Fig.~\ref{fig:reso} the real and imaginary parts of the $\bar K^*N \rightarrow \bar K^* N$ transition amplitude are displayed for different isospin sectors as functions of the center-of-mass energy $P_0$ for a total momentum $\vec{P}=0$. We analyze two different cases: (i) solution of the on-shell Bethe-Salpeter equation in free space (dashed lines), (ii) in-medium amplitude at normal nuclear matter density $\rho_0=0.17 \ {\rm fm}^{-3}$, including Pauli blocking effects on the nucleon and the $\bar K^*$ self-energy in a self-consistent manner (solid lines).

From the resonant states that are dynamically  generated in free space within the range of energies explored, two of them couple strongly to ${\bar K}^* N$, the $I=0$ $\Lambda(1783)$ and the $I=1$ $\Sigma(1830)$ states. They
are clearly visible in the corresponding isospin transition amplitude displayed in Fig.~\ref{fig:reso}.  As previously discussed in the free-space model developed in Ref.~\cite{Oset:2009vf}, there is one $I=0$ state in the PDG  with $J^P=1/2^-$, the $\Lambda(1800)$, remarkably close to the energy of one of the dynamically generated states. The spin partner is, however, absent in the PDG or corresponds to the $J^P=3/2^-$ $\Lambda(1690)$, although this would imply a large breaking of the spin degeneracy implicit in the unitary hidden-gauge model. In the $I=1$ sector, the obtained $\Sigma(1830)$ state was associated to the $J^P=1/2^-$ PDG state $\Sigma(1750) $\cite{Oset:2009vf}.
The widths of both states are, however, smaller than the ones measured experimentally.  We recall that
the main source of imaginary part of the model comes from convoluting the free loop function with the mass distributions for $\rho$ and $\bar K^*$ mesons. The inclusion of pseudoscalar baryon decays would make
the widths larger and, therefore, closer to the experimental ones.

Medium effects on the transition amplitudes come from two sources. On the one hand, we incorporate Pauli blocking on nucleons, an effect which cuts phase-space in the unitarized amplitude and pushes the resonances to higher energies. On the other hand, we include the attractive $\bar{K}^*$ self-energy in the $\bar K^*N$ intermediate states of the loop function (see later on in this section), moving the resonances back to lower energies. As a consequence, the resonances stay close to their free position. A similar behaviour was previously observed for the $\Lambda(1405)$ in the pseudoscalar-baryon sector \cite{Lutz1,Ramos:1999ku,Tolos:2000fj,Tolos:2002ud,Tolos:2006ny}.
The width of these resonances in matter, however, increases substantially due to the opening of new decay channels.
Note that the self-consistent  $\bar K^* N$ effective interaction also incorporates in the loop function of the intermediate $\bar K^*N$ states the self-energy from the $\bar K^* \rightarrow \bar K \pi$
decay mechanism, shown in Fig.~\ref{fig:kskpi}. This
implements new nucleon absorption possibilities,
such as $\bar K^* N \to \pi \bar K N$ or $\bar K^* NN \rightarrow \bar K N N$.

\begin{figure}[t]
\begin{center}
\includegraphics[width=0.5\textwidth,]{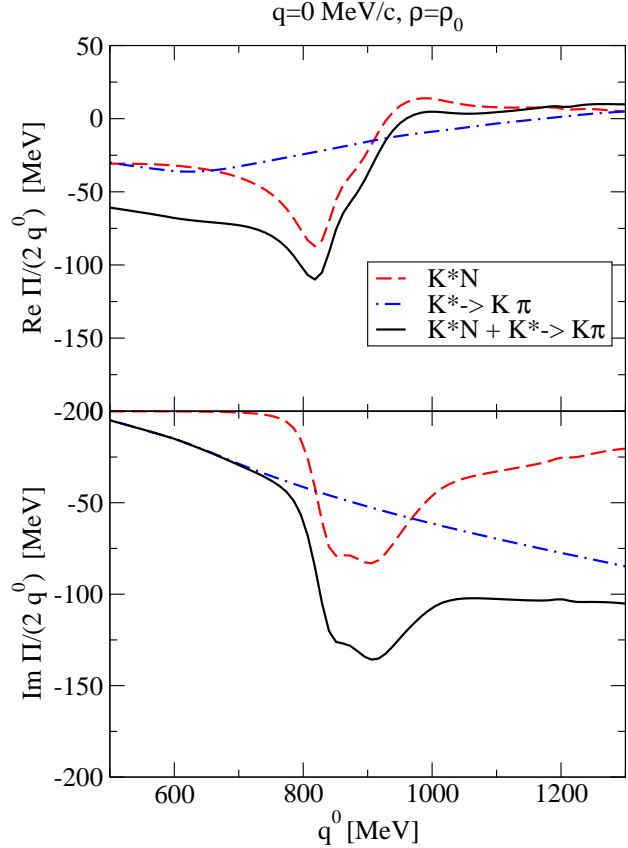}
\caption{ Real and imaginary parts of the $\bar K^*$ self-energy
as functions of the meson energy $q^0$ for zero momentum and normal
saturation density $\rho_0$ showing the different contributions: (i) self-consistent calculation of the $\bar K^* N$ interaction (dashed lines), (ii) self-energy coming from $\bar K^* \rightarrow \bar K \pi$  decay (dot-dashed lines), and (iii) combined self-energy from both previous sources (solid lines).}
\label{fig:self0}
\end{center}
\end{figure}

We show next in Fig.~\ref{fig:self0} the $\bar K^*$ self-energy as a function of the $\bar K^*$ energy $q^0$ for zero momentum at normal nuclear matter density. We display the contribution to the self-energy coming from the self-consistent calculation of the $\bar K^* N$ effective interaction (dashed lines) and the self-energy from the $\bar K^* \rightarrow \bar K \pi$ decay mechanism (dot-dashed lines), together with the combined result from both sources (solid lines).

For $\bar K^*$ energies around 800-900 MeV we observe an enhancement of the width together with some structures in the real part of the self-energy. This comes from the coupling of the $\bar K^*$ to the dynamically generated $\Lambda(1783) N^{-1}$ and  $\Sigma(1830) N^{-1}$ excitations, which dominate the behavior of the $\bar K^*$ self-energy in this energy region. However, at lower energies where the $\bar K^* N\to V B$ channels
are closed, or as energy increases,
the width of the $\bar K^*$ is governed by the $\bar K \pi$ decay mechanism in dense matter.
At the
$\bar K^*$ mass, the $\bar K^*$ feels a moderately attractive optical potential and
acquires a width of $260$ MeV, which is about 5 times its width in vacuum.
\begin{figure}[t]
\begin{center}
\includegraphics[width=0.8\textwidth]{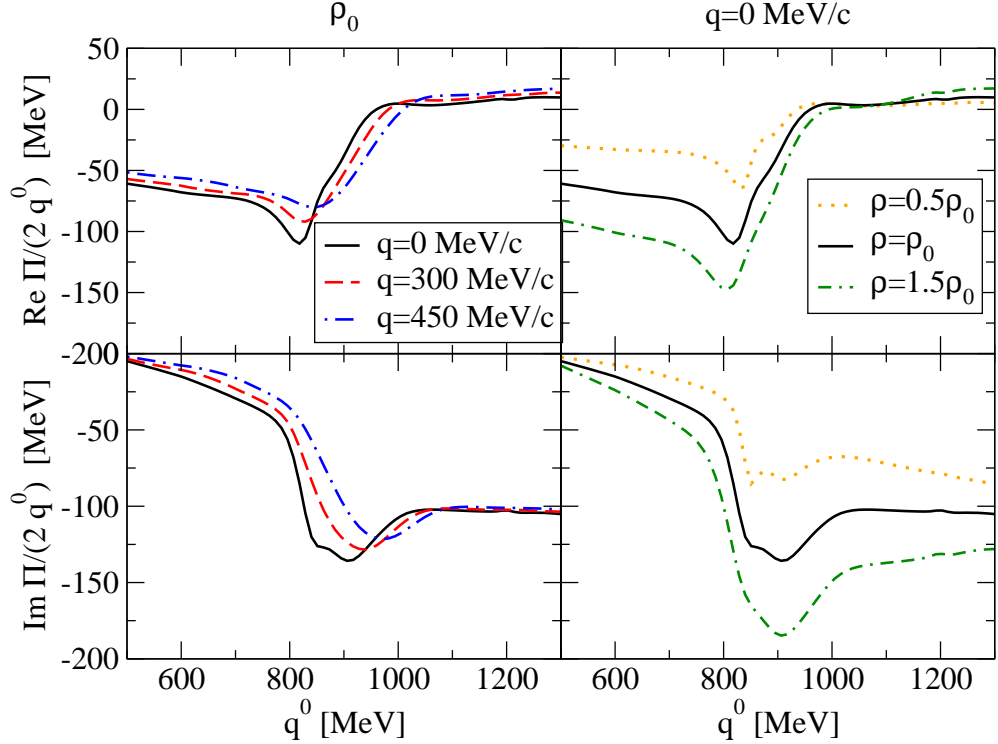}
\caption{ The $\bar K^*$ self-energy as a function of the meson energy $q^0$ for different momenta and densities.}
\label{fig:selfrq}
\end{center}
\end{figure}

The behaviour of the full $\bar K^*$ self-energy with density and momenta is analyzed in Fig.~\ref{fig:selfrq}, where we display the real and imaginary parts of the self-energy at $\rho_0$ for different momenta (left panels) and at $q=0 \ {\rm MeV/c}$ for different densities (right panels). We observe that the contribution of the resonant-hole states to the self-energy is shifted to higher energies as we move up in momenta, and it increases with density. In general, there is a systematic growth of the  imaginary part of the self-energy with the available phase space.

\begin{figure}[t]
\begin{center}
\includegraphics[width=0.5\textwidth]{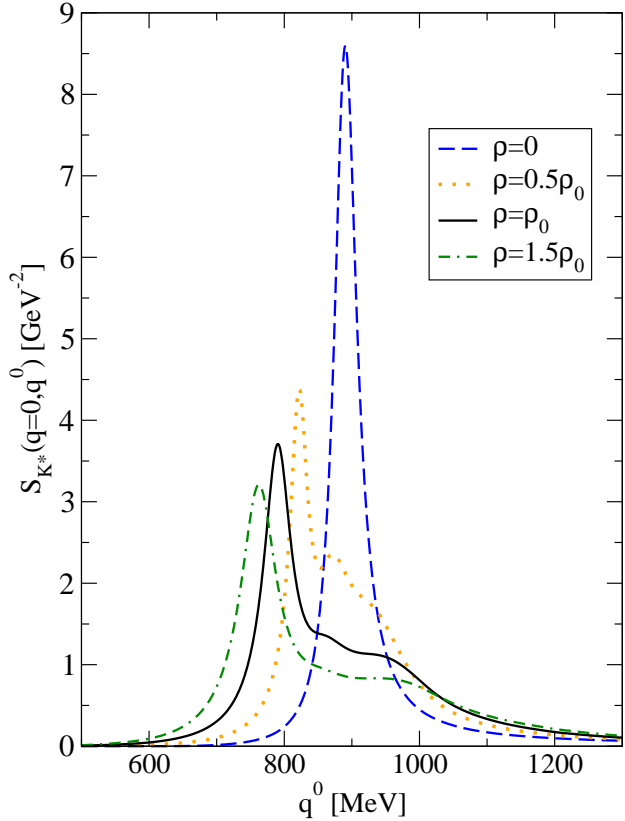}
\caption{ The $\bar K^*$ spectral function as function of the meson
  energy $q^0$ for different densities and zero momentum.}
\label{fig:spec}
\end{center}
\end{figure}

These results are better visualized in the $\bar K^*$ meson spectral function,
which is displayed in Fig.~\ref{fig:spec} as a function of the meson energy
$q^0$, for zero momentum and different densities up to $1.5\rho_0$. The dashed
line refers to the calculation in free space, where only the $\bar K \pi$ decay
channel contributes, while the other three lines correspond to fully
self-consistent calculations, which also incorporate the process $\bar K^*
\rightarrow \bar K \pi$ in the medium.

We observe a rather pronounced  peak at the quasiparticle energy
\begin{equation} \omega_{\rm qp}(\vec{q}=0\,)^2=m^2+ {\rm
Re}\Pi(\omega_{\rm qp}(\vec{q}=0\,),\vec{q}=0\,) \ , \label{eq:qp}
\end{equation}
which moves to lower energies with respect to the free $\bar K^*$ mass
position as density increases. The $\Lambda(1783) N^{-1}$ and  $\Sigma(1830) N^{-1}$ excitations
are also clearly visible on the right-hand side of the quasiparticle peak.
In spite of the fact that the real part of the optical potential, ${\rm
Re}\Pi(m_{\bar K^*},\vec{q}=0\,)/(2 m_{\bar K^*})$, acquires a moderate value
of $-50$ MeV at
$\rho_0$ (see Fig. \ref{fig:self0}), interferences with the resonant-hole modes, which
appear at energies close to the
${\bar K}^*$ mass, push the quasiparticle peak to a substantially lower energy.
Note, however, that the properties of the quasiparticle peak will be
affected by the coupling of the $\bar K^*$ to other subthreshold excitations,
such as ${\bar K}N N^{-1}$, $\pi Y N^{-1}$, $Y N^{-1}$,..., not accounted for in
the present model. The peak will be wider, less
pronounced, and, through self-consistency, might even be pushed back to higher
energies, mixing with the resonant-hole
modes that will also become more extended.
Density
effects result in a dilution and merging of those resonant-hole states,
together with a general broadening of the spectral function  due to the
increase of collisional and absorption processes.

From the above considerations, a clear conclusion
of the present work is
that the ${\bar K^*}$
experiments a tremendous increase of width in matter, a fact that can be tested
via a transparency ratio experiment, as discussed in section VI.

\section{Overview of the method, critical discussion and future developments}

In this section we wish to comment on the method used, discussing missing terms, with an estimation of their size, as well as other issues where improvements can be done in the future.

The first thing that deserves some explanation is why we distinguish between the $\bar K^*$ self-energy from the decay into $\bar K \pi$ and the one coming from the $\bar K^* N$ interaction. The main answer is because they provide different sources of inelastic $\bar K^* N$ scattering which sum incoherently in the $\bar K^*$ width. Indeed, when studying the $s$-wave $\bar K^* N$ interaction we have the coupled channels $\bar K^*N$, $\omega \Lambda$, $\rho \Sigma$,
$\phi \Lambda$ and $K^* \Xi$ for $I=0$,  while for $I=1$ we consider
$\bar K^*N$, $\rho \Lambda$, $\rho \Sigma$, $\omega \Sigma$, $K^* \Xi$ and
$\phi \Sigma$ (see Section \ref{freespace}). The inelastic channels are all of the vector meson-baryon type. Instead, for the in-medium corrections to the decay width, represented by the diagram on the right of Fig.~\ref{fig:1}, we have a pseudoscalar meson and a baryon in the inelastic channels, since the processes induced are of the type $\bar K^* N \rightarrow \bar K N$ when the $\pi$ excites a $ph$, or $\bar K^*N \rightarrow \pi \Lambda, \pi \Sigma$ when the ${\bar K}$ couples to a $Yh$ excitation.
 Incidentally, we could have both $ph$ and $Y h$ excitations simultaneously and this would contribute to the absorption channels $\bar K^* N N \rightarrow \Lambda N, \Sigma N$. All these channels add incoherently to the width of the $\bar K^*$ since they correspond to different final states.

The evaluation of the $\bar K^* N \rightarrow \bar K N$ transition amplitude from the diagram of Fig.~\ref{fig:1} (right) corresponds to having it mediated by pion exchange, which is a fair assumption. This is different from the way we evaluate the transition amplitudes
$\bar K^* N \rightarrow \bar K^* N, V Y$ between coupled vector-baryon channels. In this case, we rely on the hidden local gauge approach to derive a tree-level transition potential mediated by vector exchange, and use it in a Bethe-Salpeter coupled-channel equation to evaluate the elastic $\bar K^* N \rightarrow \bar K^* N$ amplitude. This amplitude is unitary and through its imaginary part, via the optical theorem, one makes connection with the vector-baryon inelastic channels. Ideally, one might like to put together the vector-baryon and pseudoscalar-baryon states in the coupled-channel unitary equation. However, at the energies of interest, transitions of the type $\bar K^* N \rightarrow \bar K N$ mediated by pion exchange may allow for this pion to be placed on its mass shell. This forces one to keep track of the proper analytical cuts contributing to the imaginary part of the diagram of Fig.~\ref{fig:1} (right), and hence to the $\bar K^*$ width, making the iterative problem technically more complicated. Note also that because of the VPP and PBB vertices one is mixing partial waves in the $\bar K^* N \rightarrow \bar K N$ transition. This problem occurs also in other reactions, such as in $\rho \rho \rightarrow \rho \rho$, where the $\rho$ can decay into $\pi \pi$, which generates the inelastic channel $\rho \rho \rightarrow \pi \pi$ mediated by $\pi$ exchange (see \cite{molinavec,gengvec}).

A technical solution was provided in these latter works, consisting of calculating the box diagram $\rho \rho \rightarrow \pi \pi \rightarrow \rho \rho$, with $\rho\rho \to \pi\pi$ mediated by pion exchange, in a field-theoretical way taking into account all the cuts properly. The resulting $\rho \rho \rightarrow \rho \rho$ term was added to the $\rho \rho \rightarrow \rho \rho$ potential coming from vector-meson exchange, and the combination was then used as the kernel of the Bethe-Salpeter equation. A welcome feature of this calculation is that the real part of the box diagram is very small and does not change the energies of the dynamically generated resonances. The imaginary part, however, has the effect of inducing an additional width to these resonances, which then stand as largely vector-vector molecules which decay into pseudoscalar-pseudoscalar pairs. The same occurs in the generalization to $SU(3)$ of the vector-vector interaction studied in \cite{gengvec}.

The model of the vector-baryon interaction used here produces quite narrow dynamically generated states compared to the their experimental counterparts, which was precisely attributed in \cite{Sarkar:2009kx,Oset:2009vf} to, essentially, the neglect of pseudoscalar-baryon channels. It is clear that including these channels as box diagrams, in complete analogy with what has been done in the vector-vector case, would be most welcome and will be done in the near future. In the mean time, we can give here an estimation of what the effect of such improvement would be. We recall, from our discussion of Fig.~\ref{fig:spec}, that the $\bar K^*$-induced resonance-hole excitation strength appears as two small bumps in the middle of the wide $\bar K^*$ spectral function. Having these resonances with a larger width from the incorporation of the pseudoscalar-baryon box diagrams, would only lead to a further redistribution of the strength of the spectral function, softening even more the already soft effects of these resonance-hole excitations, without altering our main conclusion about the large width of the $\bar K^*$ in nuclei from its many decay channels.

There is another issue worth mentioning. In the approach of \cite{molinavec,gengvec,Sarkar:2009kx,Oset:2009vf}, the interaction of vectors with vectors or baryons has been studied at low energies, assuming that the vector three momenta are small compared to the vector mass. Under these circumstances, the approach becomes technically easy and, among other simplifications, one can approximate by a constant the propagator of the vector meson exchanged between the hadronic components. This comes because
the momentum transferred $\vec{q}$ is negligible versus the vector meson mass $M_V$
($\vec{q}\,^2/M_V^2\ll 1$), rendering the interaction to be of contact type.
This is implicit in the chiral lagrangians, which can also be deduced from the hidden gauge approach making this approximation. While this assumption is unquestionable in the limit $\vec{q}\,^2 \rightarrow 0$, there is the fair question of exploring its limits of applicability. This issue is discussed in detail in Section 2.2 of \cite{Sarkar:2009kx}, to where we convey the reader for more details. There, the role of the $t$-channel exchange of vector mesons, as well as the $u$-channel processes with baryon exchange, are thoroughly discussed and warnings are given about danger of misuse when dealing with low-energy lying channels. The conclusion of that section is that both the local approximation of the $t$-exchange interaction and the neglect of $u$-channel exchange for the $s$-waves are well justified approximations for the range of energies studied in \cite{Sarkar:2009kx,Oset:2009vf}, which is also the range of relevance for the present work.

A further mechanism when dealing with vector-baryon interaction is the exchange of pseudoscalar mesons rather than vector mesons within the vector-baryon coupled channels. This involves a vector-vector-pseudoscalar vertex, of anomalous type. This mechanism has been studied in \cite{pedro}, where the $\rho \Delta \rightarrow \omega \Delta$ transition was studied in detail. It was shown that the corresponding box diagram, $\rho \Delta \rightarrow \omega \Delta \rightarrow \rho \Delta$, was negligible compared to the hidden gauge local term of the $\rho \Delta \rightarrow \rho \Delta$ interaction. The same conclusion was reached in \cite{Sarkar:2009kx}, where the $\rho \Delta \rightarrow \omega N \rightarrow \rho \Delta$ box diagram was explicitly evaluated. These findings followed similar conclusions about the contribution of these anomalous terms in the vector-vector interaction \cite{molinavec,gengvec}.

Another aspect worth discussing, since it has not been addressed in previous works
 \cite{pedro,Sarkar:2009kx,Oset:2009vf} concerns the use, in the hidden gauge method, of only the vector-type coupling ($\gamma^{\mu}$) of the vector mesons to the baryons, Eq.~(\ref{lagr82}), whereas the tensor-type coupling ($\sigma^{\mu \nu}$), known to be sizable for the $\rho NN$ case, is neglected. The $\rho N N$ coupling reads \cite{palomar}
\begin{eqnarray}
- {\rm i} t_{\rho N N}= {\rm i} \left( G^{V} \gamma^{\mu}+\frac{G^{T}}{2 \ {\rm i} \ M_N} \ \sigma^{\mu \nu} \ q_{\nu} \right) \epsilon^*_{\mu} \tau^a \ ,
\end{eqnarray}
with $G^{V}=2.9 \pm 0.3$ and $G^{T}=18 \pm 2$, where $q_{\mu}$ is the $\rho$ momentum and $\epsilon_{\mu}$ its polarization vector. The non-relativistic reduction of this vertex reads \cite{palomar}
\begin{eqnarray}
- {\rm i} t_{\rho N N}= {\rm i} \left( G^{E} \epsilon^{0} - \frac{G^{M}}{2 \ M_N} \ ( \vec{\sigma} \times \vec{q}\,) \vec{\epsilon} \right)  \tau^a \ ,
\label{eqvector}
\end{eqnarray}
with $G^E=G^{V}$, $G^{M}=G^{V}+G^{T}$.

\begin{figure}[t]
\begin{center}
\includegraphics[width=0.5\textwidth]{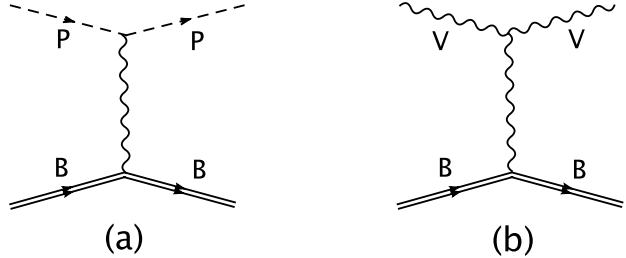}
\caption{Diagrams contributing to the pseudoscalar-baryon (a) or vector-
baryon (b) interaction via the exchange of a vector meson.}
\label{fig:art}
\end{center}
\end{figure}

This tensor coupling in the case of the $\rho$ is known to be relevant when studying the nucleon-nucleon tensor force and the spin-isospin excitation modes in nuclei \cite{weiserep}. Yet, for the case of the vector-baryon interaction the situation is quite different. Indeed, as shown in \cite{Sarkar:2009kx,Oset:2009vf}, the three vector vertex, $VVV$ in Fig.~\ref{fig:art}(b) is the same as the $PPV$ of Fig.~\ref{fig:art}(a) for small values of the three-momentum of the vector mesons. The problem of neglecting the tensor-type coupling in the lower $VBB$ vertex  is therefore also present in the study of the interaction of pseudoscalar mesons with baryons. Yet, this interaction has been traditionally tackled by means of chiral lagrangians \cite{ecker,ulfrep}, which stem from diagram of Fig.~\ref{fig:art}(a), calculated within the hidden gauge approach, neglecting $\vec{q}\,^2/M_V^2$ and considering only the $\gamma^{\mu}$ coupling of the $VBB$ vertex. The success of the chiral unitary approach in the meson-baryon interaction is suggesting that there must be a good justification to neglect the tensor coupling in this type of problems. Since this has not been explicitly exposed in earlier works, we take advantage here to make estimates of the relevance of this term. Indeed, let's consider the transition $K^{* -} p \rightarrow K^{* -} p$ represented by the diagram of Fig.~\ref{fig:art}(b). For small three-momenta of the vector mesons, the $VVV$ vertex behaves as the $PPV$ vertex of diagram of  Fig.~\ref{fig:art}(a), namely  $(k+k')_{\mu} \epsilon^{\mu}$, with $k,k'$ the momenta of the kaons and $\epsilon^{\mu}$ the vector polarization. When contracting this vertex with the $BBV$ one of Eq.~(\ref{eqvector}), we get two contributions
\begin{eqnarray}
G^E(k^0+k^{'0}) \ \ \ \ , \ \ \ \ G^{M} \left( \frac{\vec{\sigma} \times \vec{q}\, }{2 M_N} \right) (\vec{k}+\vec{k'}) \ ,
\end{eqnarray}
with $\vec{q}=\vec{k}-\vec{k'}$, by means of which we obtain, close to the  $K^{*-}p$ threshold, the terms $2G^EM_V$ versus $\frac{G^M}{M_N} \vec{\sigma} (\vec{k} \times \vec{k'})$. 

First, note that these two terms belong to different partial waves and they do not add coherently. The term with $G^M$ also vanishes when we evaluate the $t \rho$ optical potential in nuclear matter for the $\bar K^*$, where $t$ stands for the forward $K^{*-} p \rightarrow K^{*-} p$ amplitude. The contribution of the magnetic term to the $\bar K^*$ potential, or to the imaginary part of the $\bar K^* N \rightarrow \bar K^* N$ amplitude comes from the angled averaged amplitude squared, $\int d\Omega |T|^2$. We also recall that 
the $K^{* -} K^{* -} \rho$ and $K^{* -} K^{* -} \omega$ vertices are the same
and the fact that
the $\gamma^{\mu}$ part of the $\omega p p $ vertex is three times bigger than the corresponding one of the $\rho p p$ vertex, while the tensor $\omega p p $ one is negligible.  Taking all these considerations into account
and assuming $|\vec{k}|=|\vec{k'}| \approx$ 300 MeV, 
when evaluating ${\rm Im} t_{\bar K^*p \rightarrow \bar K^* p}$, or the cross section, the relative contribution of the magnetic $\rho$ term to the 
sum of the $\rho$ and $\omega$ electric terms comes out of the order of 0.7 \% (2 \% for  $|\vec{k}|=|\vec{k'}| \approx$ 500 MeV). This exercise justifies a posteriori why the magnetic $\rho N N$ coupling has been systematically neglected in the chiral dynamics studies of the meson-baryon or vector-baryon interaction, and, for our purposes, why we neglect it here too.

\section{Nuclear transparency in the $\gamma ~A \to   K^+ ~K^{*-}~ A'$ reaction}

  In this section we make a qualitative evaluation of the nuclear transparency ratio by comparing the cross sections of the photoproduction reaction $\gamma A \to K^+ K^{*-} A'$ in different nuclei, and tracing them to the in medium width of the $K^{*-}$. The idea is that the survival probability is an exponential function of the integral of the in-medium width, and hence very sensitive to this magnitude \cite{Hernandez:1992rv}.

We write the
nuclear transparency ratio as
\begin{equation}
\tilde{T}_{A} = \frac{\sigma_{\gamma A \to K^+ ~K^{*-}~ A'}}{A \,\sigma_{\gamma N \to K^+ ~K^{*-}~N}} \ ,
\end{equation}
{\it i.e.} the ratio of the nuclear $K^{*-}$-photoproduction cross section
divided by $A$ times the same quantity on a free nucleon.
The value of $\tilde{T}_A$ describes
the loss of flux of $K^{*-}$ mesons in the nucleus and is related to the
absorptive part of the $K^{*-}$-nucleus optical potential and thus to the
$K^{*-}$ width in the nuclear medium.  This method
has been already proven to be very efficient in the study of the in-medium
properties of the vector mesons~\cite{magas,Muhlich:2005kf,Kaskulov:2006zc},
hyperons~\cite{Kaskulov:2006nm} and antiprotons \cite{Hernandez:1992rv}. In Ref.~\cite{muhlich2,:2008xy}
the transparency ratio has been
already used to determine the width of the $\omega$-meson in finite nuclei
using a BUU transport approach.

We have done  calculations for a vast sample of nuclear targets:
${}^{12}_6$C, ${}^{14}_{7}$N,
${}^{23}_{11}$Na,  ${}^{27}_{13}$Al, ${}^{28}_{14}$Si,
${}^{35}_{17}$Cl,   ${}^{32}_{16}$S,  ${}^{40}_{18}$Ar, ${}^{50}_{24}$Cr,
${}^{56}_{26}$Fe, ${}^{65}_{29}$Cu,  ${}^{89}_{39}$Y, 
${}^{110}_{48}$Cd,  ${}^{152}_{62}$Sm,  ${}^{207}_{82}$Pb,
${}^{238}_{92}$U.

 In the following, we evaluate the ratio between the nuclear
cross sections in
heavy nuclei and a light one, for instance  $^{12}$C, since in
this way, many other nuclear effects not related to the
absorption of the $K^{*-}$ cancel in the
ratio~\cite{magas}. We call this ratio $T_A$,
\begin{equation}
T_{A} = \frac{\tilde{T}_{A}}{\tilde{T}_{^{12}C}} \ ,
\end{equation}
and, by construction, it is normalized to unity for $^{12}$C.

We obtain the nuclear transparency ratio taking an eikonal (or Glauber) approximation in the evaluation of
the distortion factor associated
to $K^{*-}$ absorption.
In this framework,
the propagation of the $K^{*-}$ meson in its way out of the nucleus is implemented by means of
the exponential factor for the survival probability
accounting for the loss of flux per unit
length. This simple but rather reliable method allows us to get
an accurate result for the integrated cross sections.

We proceed as follows: let $\Pi_{K^{*-}}$ be
the $K^{*-}$ self-energy in the nuclear medium
 as a function
of the nuclear density, $\rho(r)$. We then have
\begin{equation}
\frac{\Gamma_{K^{*-}}}{2} = - \frac{\mbox{Im}\Pi_{K^{*-}}}{2 E_{K^{*-}}};
\qquad
\Gamma_{K^{*-}} \equiv\frac{d{P}}{dt} \ ,
\end{equation}
where  $P$ is the probability of $K^{*-}$ interaction in the
nucleus, including $K^{*-}$ quasi-elastic collisions and
absorption channels. There is some problem when dealing with the free part of the $K^{*-}$ self-energy. Indeed, if the $K^{*-}$ decays inside the nucleus into $\bar{K} \pi$, the $\bar{K}$ or the $\pi$ will also be absorbed with great probability or undergo a quasi-elastic collisions, such that the $\bar{K} \pi$ invariant mass will no longer be the one of the $K^{*-}$.
Thus, it is adequate to remove these events.  Yet, if the decay occurs at the surface neither the $\bar{K}$ nor the $\pi$ will be absorbed and an experimentalist will reconstruct the $K^{*-}$ invariant mass from these two particles, in which case they should not be removed from the flux. The part of $\mbox{Im}\Pi_{K^{*-}}$ due to quasi-elastic collisions $K^{*-} N  \to K^{*-} N $ should not be taken into account in the distortion either, since  the $K^{*-}$ does not disappear from the flux
in these processes. Yet, this part is small at low energies and we disregard this detail in the present qualitative estimate. In view of all this, we have taken the following  approximate choice for $\mbox{Im}\Pi_{K^{*-}}$ in our estimate of  $T_A$
\begin{equation}
\mathrm{Im}\,\Pi_{\bar K^*}=\left\{ \begin{array}{ll}
(-2.5\times 10^5 + 0.4\times 10^5) \rho(r)/\rho_0~\mathrm{MeV^2}  -0.4\times10^5~\mathrm{MeV^2}   & \mathrm{for} \hspace{0.1cm} r < 0.8\,R \ , \\
(-2.5\times 10^5 + 0.4\times 10^5 )\rho(r)/\rho_0 ~ \mathrm{MeV^2}
 & \mathrm{for}  \hspace{0.1cm} r\geq 0.8\,R \ ,
\end{array} \right.
\label{eq:impi}
\end{equation}
where $R$ is the nuclear radius.
The choice of Eq. (\ref{eq:impi}) is justified by the fact that $\mathrm{Im} \Pi_{\bar K^*}=-2.5\times 10^5$ MeV$^2$ at $q^0=m_{\bar{K^*}}$ and $\vec{q}=0$ MeV/c at $\rho=\rho_0$ (see Fig.~\ref{fig:selfrq}). This value also contains the contribution from the free decay, $\mathrm{Im}\Pi^{0}_{\bar K^*} \simeq -0.4\times 10^5$ MeV$^2$, which does not depend on $\rho$. It therefore
needs to be subtracted from the term that implements the linear $\rho$ dependence, and added as a constant value if $r < 0.8\,R $. Moreover, when the  $\bar{K}^*\to \bar{K}\pi$ decay
takes place in the surface of the nucleus, $r\geq 0.8\,R $, we remove
$\mathrm{Im}\Pi^{0}_{\bar K^*}$
 since experimentally the $\bar{K}\pi$ system will be reconstructed as a $\bar{K}^*$.

The probability of loss of
flux per unit length is given by:
\begin{equation}
\frac{dP}{dl}=\frac{dP}{v\,dt}
=\frac{dP}{\displaystyle \frac{|\vec{p}_{K^{*-}}|}{E_{K^{*-}}} dt}
=-\frac{\mbox{Im}\Pi_{K^{*-}}}{|\vec{p}_{K^{*-}}|} \ ,
\end{equation}
and the corresponding survival probability is determined from
\begin{equation}
\label{EF}
 \exp\left\{
\int_0^{\infty} dl
 \frac{\mbox{Im}\Pi_{K^{*-}}(\rho(\vec{r}\, '))}{|\vec{p}_{K^{*-}}|}\right\},
\end{equation}

\noindent where
 $\vec{r}\,'=\vec{r}+l\displaystyle\frac{\vec p_{K^{*-}}}{|\vec p_{K^{*-}}|}$
 with  $\vec{r}$ being the $K^{*-}$ production point inside the
nucleus.

With all these ingredients and taking into account the standard expression for $K^{*-}$ production in the nucleus prior to the consideration of the eikonal distortion, the cross section for the $\gamma ~A \to   K^+ ~K^{*-} ~ A'$ reaction is obtained from
\begin{eqnarray}
\sigma_{\gamma A \to K^+ K^{*-} A'}&=&\frac{M^2}{4 (s-M^2)}\frac{1}{(2\pi)^4}\int d^3 r \rho(r)\int^{E_2^{\mathrm{max}}}_{m_{K^{*-}}} p_2 dE_2 \int^{E_3^{\mathrm{max}}}_{m_{K^+}} dE_3 \int^1_{-1} d\mathrm{cos}\theta_2\nonumber\\&\times&\int^{2\pi}_0 d\phi_2\frac{1}{\vert\vec{p}_\gamma-\vec{p}_2\vert}\theta(1-A^2) \theta(E_\gamma+M-E_2-E_3)\vert T \vert^2\nonumber\\&\times&
\exp\left\{\int^{2.5 R}_0  dl\,\mathrm{Im}\Pi(\rho(\vec{r}\,'))/p_2\right\}\ ,
\end{eqnarray}
with
\begin{eqnarray}
A\equiv \mathrm{cos} \theta_3 =\frac{1}{2\vert \vec{p}_\gamma-\vec{p}_2\vert p_3}\lbrace M^2+\vert \vec{p}_\gamma-\vec{p}_2\vert^2+\vec{p}\,^2_3-(E_\gamma+M-E_2-E_3)^2\rbrace
\end{eqnarray}
and
\begin{eqnarray}
E^{\mathrm{max}}_2=\frac{s+m^2_{K^{*-}}-(M+m_{K^+})^2}{2\sqrt{s}}\nonumber\\
E^{\mathrm{max}}_3=\frac{s+m^2_{K^+}-(M+m_{K^{*-}})^2}{2\sqrt{s}} \ ,
\end{eqnarray}
where ($E_2,\vec{p}_2$) and ($E_3,\vec{p}_3$) are the four-momenta of the
 $K^{*-}$ and $K^+$, respectively, in the frame of the nucleon at rest, and
 $E_\gamma$ is the energy of the photon in this frame. Here, $M$ is the mass of
 the nucleon while $m_{K^{*-}}$ and $m_{K^+}$ are the masses of the $K^{*-}$
 and $K^+$ mesons, respectively. The value of $E^{\mathrm{max}}_2$  has been
 calculated when the particles 1 plus 3 go together (or 1 plus 2 in the case of
 $E^{\mathrm{max}}_3$).  Since we are interested in ratios of cross sections we
 have taken $|T|^2=1$.

\begin{figure}[t]
\begin{center}
\includegraphics[width=0.8\textwidth]{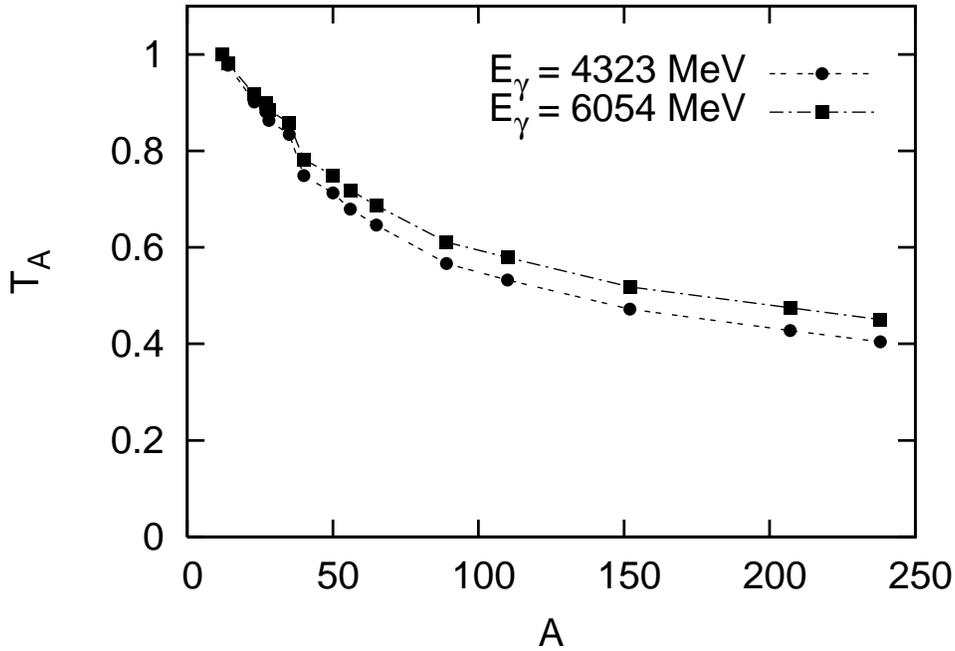}
\caption{Transparency ratio}
\label{fig:trans}
\end{center}
\end{figure}

The results can be seen in Fig. \ref{fig:trans}, where the transparency ratio
has been plotted for two different energies in the center of mass reference
system $\sqrt{s}=3$ GeV and $3.5$ GeV, which are equivalent to energies of the
photon in the lab frame of $4.3$ MeV and $6$ MeV, respectively. We observe a
very strong attenuation of the $\bar{K}^*$ production process due to the decay
or absorption channels $\bar{K}^*\to \bar{K}\pi$ and $\bar{K}^*N\to V Y$ with
increasing nuclear-mass number $A$. This is due to the larger path that the
$\bar{K}^*$ has to follow before it leaves the nucleus, having then more
chances to decay or get absorbed.

\section{Conclusions}

We have studied the properties of $\bar K^*$ mesons in symmetric
nuclear matter within a self-consistent coupled-channel
unitary approach using hidden-gauge local symmetry.
The corresponding in-medium solution incorporates Pauli blocking
effects and the $\bar K^*$ meson self-energy in a
self-consistent manner. In particular, we have analyzed the behavior
of dynamically-generated baryonic resonances in the nuclear medium and
their influence in the self-energy and, hence, the spectral
function of the $\bar K^*$ mesons.

We have found a moderate attractive optical potential for the ${\bar K}^*$
of the order of $-50$ MeV at normal nuclear matter density. The corresponding
quasiparticle peak in the spectral function might
not be easily distinguished due its merging with other excitations,
apart from the fact that changes of mass are always very difficult to
determine in experiments \cite{Nanova:2010sy}. More remarkable are the changes in
the width, which can be more easily addressed by means of transparency ratios
in different reactions. At normal nuclear matter density the $\bar{K}^*$ width is
found to be about $260$ MeV, five times larger than its free width.
This spectacular increase is much bigger than the width of the $\rho$ meson in
matter, evaluated theoretically in
\cite{Rapp:1997fs,Peters:1997va,Urban:1999im,Cabrera:2000dx} or measured recently
\cite{Arnaldi:2006jq, Arnaldi:2008er,Wood:2008ee,Djalali:2008zza}.

We have made an estimation of the transparency ratios in the $\gamma A\to
K^+\bar{K}^* A'$ reaction and found a substantial reduction from unity of that
magnitude, which should be easier to observe experimentally. Other reactions
like the $K^- A\to K^{*-} A'$ should also be good tools to investigate these
important changes linked to the strong $\bar{K}^*$ interaction with the nuclear
medium \cite{Djalali}.

\section{Acknowledgments}
 L.T. wishes to acknowledge support from the Rosalind Franklin Programme of the
 University of Groningen (The Netherlands) and the Helmholtz
 International Center for FAIR within the framework of the LOEWE
 program by the State of Hesse (Germany). This work is partly supported by  projects FIS2006-03438, FIS2008-01661 from the Ministerio de Ciencia e Innovaci\'on (Spain), by the Generalitat Valenciana in the program Prometeo and
 by the Ge\-ne\-ra\-li\-tat de Catalunya contract 2009SGR-1289.
 This research is part of the European
 Community-Research Infrastructure Integrating Activity ``Study of
 Strongly Interacting Matter'' (acronym HadronPhysics2, Grant
 Agreement n. 227431) and of the EU Human Resources and Mobility
 Activity ``FLAVIAnet'' (contract number MRTN--CT--2006--035482),
 under the Seventh Framework Programme of EU.


\end{document}